\documentclass[lettersize,journal]{IEEEtran}
\usepackage[nocompress]{cite}
\usepackage{amsmath,amsfonts}
\usepackage{algorithmic}
\usepackage{array}
\usepackage[caption=false,font=footnotesize,labelfont=rm,textfont=rm]{subfig}
\usepackage{textcomp}
\usepackage{stfloats}
\usepackage{url}
\usepackage{threeparttable}
\usepackage{verbatim}
\usepackage{graphicx}
\def\BibTeX{{\rm B\kern-.05em{\sc i\kern-.025em b}\kern-.08em
    T\kern-.1667em\lower.7ex\hbox{E}\kern-.125emX}}
\usepackage{balance}

\usepackage{tikz}
\usepackage{amsmath}
\usepackage[english]{babel}
\usepackage{blindtext}
\usepackage[ruled,vlined,linesnumbered]{algorithm2e}
\usepackage{multirow}
\usepackage{makecell} 
\usepackage{booktabs} 
\begin{document}
\title{FluxShard: Motion-Aware Feature Cache Reuse for Collaborative Video Analytics in Mobile Edge Computing}
\author{Xiuxian Guan, Zongyuan Zhang, Zheng Lin, Zekai Sun, Tianyang Duan, Zihan Fang, Rui Wang, Heming Cui, Wei Ni,~\IEEEmembership{Fellow,~IEEE}, Jun Luo,~\IEEEmembership{Fellow,~IEEE}, and Yuanwei Liu,~\IEEEmembership{Fellow,~IEEE}
\thanks{Xiuxian Guan, Zongyuan Zhang, Zekai Sun, Tianyang Duan, and Heming Cui are with the Department of Computer Science, The University of Hong Kong, Hong Kong, China. (email: xxguan@cs.hku.hk; zyzhang2@cs.hku.hk; zksun@cs.hku.hk; tyduan@cs.hku.hk; heming@cs.hku.hk)}
\thanks{Xiuxian Guan and Rui Wang are with the Department of Electronic and Electrical Engineering, The Southern University of Science and Technology, Shenzhen, China. (email: guanxx2021@mail.sustech.edu.cn; wang.r@sustech.edu.cn)}
\thanks{Zheng Lin and Yuanwei Liu are with the Department of Electrical and Computer Engineering, University of Hong Kong, Pok Fu Lam, Hong Kong, China (email: linzheng@eee.hku.hk; yuanwei@hku.hk).}
\thanks{Zihan Fang is with Hong Kong JC STEM Lab of Smart City and Department of Computer Science, City University of Hong Kong, Kowloon, Hong Kong SAR, China (e-mail: zihanfang3-c@my.cityu.edu.hk).}
\thanks{Wei Ni is with the School of Engineering, Edith Cowan University, Perth, WA 6027, Australia (email: wei.ni@ieee.org).}
\thanks{Jun Luo is with the College of Computing
and Data Science, Nanyang Technological University, Singapore (e-mail: junluo@ntu.edu.sg).}
}

\markboth{IEEE Transactions on Mobile Computing}%
{FluxShard: Motion-Aware Feature Cache Reuse for Collaborative Video Analytics in Mobile Edge Computing}

\maketitle

\begin{abstract}
	Caching and reusing intermediate features across consecutive frames is a common
    technique to reduce redundant computation and transmission for
	edge-cloud video analytics in mobile edge computation. Existing methods manage the cache in a fixed
	or globally shifted coordinate system, treating it as an indivisible whole.
	Under the non-uniform motion patterns of mobile scenes, this whole-scene
	granularity invalidates large portions of the cache even when most content
	has merely shifted spatially, wasting computation and bandwidth.
	The root cause is a granularity mismatch: the cache is managed per scene,
	yet motion varies per region. In this paper, we present FluxShard, a motion-aware edge-cloud video analytics system that
	uses codec-level block motion vectors (MVs) to manage feature cache
	reuse and recomputation at the granularity of individual motion regions.
	By re-indexing cached features along per-block MVs, FluxShard separates
	spatial displacement from content changes, recovering reusable
	content that whole-scene methods would otherwise discard.
	To ensure correct reuse under heterogeneous motion, the Receptive Field Alignment Principle (RFAP)
	identifies, from the input-level MV field alone, the positions that
	must be recomputed due to inconsistent spatial composition within
	receptive fields.
	To maintain cache coherence across frames, MV-guided cache remapping
	warps the entire feature cache to the current coordinate system each
	frame, sustaining a high reuse ratio over time.
	A profiling-driven dispatcher routes the remaining sparse
	workload between edge and cloud for lower latency.
	Evaluation across multiple vision tasks, dynamic video benchmarks,
	and network conditions shows that FluxShard reduces latency by
	32.6--83.8\% and energy by 14.9--64.0\% over all baselines under the prescribed accuracy budget.
\end{abstract}


\section{Introduction}

Video analytics is a core building block of intelligent edge applications in mobile edge computation: autonomous drones~\cite{s23218711,nguyen_person_2025}, embodied robots~\cite{wu_helpful_2024,han_fetchbench_2024}, augmented-reality headsets~\cite{zhang_vdo-slam_2021}, and smart surveillance cameras~\cite{jalali_robust_2025} all rely on video analytics to perceive and react to their surroundings in real time.
It uses deep neural networks (DNNs)~\cite{lin2024efficient,fang2024ic3m,yuan2024satsense,lin2026gapsl,hong2026conflict,peng2024sums,zhang2024satfed,lin2024fedsn} to process continuous video frames and produce results such as object detections, segmentation masks, and pose estimates.
Running these models directly on edge devices avoids uploading raw video and keeps latency predictable, but edge hardware offers limited compute and energy.
A common alternative is to offload frames to a powerful cloud server; however, streaming high-resolution video over a bandwidth-constrained wireless uplink introduces transmission delay that can violate the real-time requirements.
Neither naive local nor remote processing reliably meets the joint latency and accuracy demands of these applications. A natural way to reduce both transmission and computation is to exploit temporal redundancy: consecutive frames share most of their content, and this redundancy extends to intermediate feature maps. Reuse-based methods cache and reuse previously computed feature maps and recompute only the changed portions, reducing both transmitted data and computation simultaneously.

Existing reuse methods are designed primarily for fixed cameras and struggle with the complex motion patterns introduced by mobile edge devices. Some methods reuse whole intermediate feature maps only when consecutive frames exceed a global similarity threshold~\cite{spinn,coach}. Others compute per-pixel differences and recompute only positions where the difference exceeds a threshold~\cite{cbinfer,diffy,deltacnn,motiondeltacnn}. Despite their different granularities, these methods share a common limitation: they treat the cached feature maps as an indivisible whole; the non-uniform motion patterns in real-world scenes invalidate large portions of the cache even when most content has merely shifted spatially, wasting computation and bandwidth. A cache reuse mechanism that adapts to local, heterogeneous motion is therefore needed for mobile video analytics.

Our insight is that block-level motion vectors (MVs) from standard video codecs provide exactly such a per-region motion signal at no additional cost. MVs are block-level displacements that map each current-frame block to the reference frame. Because MVs capture geometric displacement rather than content change, re-indexing cached feature-map blocks along their MV trajectories recovers reusable content that whole-scene methods would otherwise discard. The remaining positions where MVs cannot account for the actual change (e.g., high difference in pixel values after alignment due to non-rigid deformation, dis-occlusion, or newly appearing content) form the \emph{recomputation set}: the minimal set of positions that must be freshly computed and transmitted.

Realizing this idea introduces two challenges.
One of the challenges is \emph{receptive field inconsistency}: even when every pixel individually matches its MV-aligned counterpart, the reused output can still be wrong. Operators like convolutions compute each output from a spatial neighborhood (the receptive field); correctly reusing a cached output requires verifying that the entire neighborhood matches, and this check at every layer and position costs as much as recomputation, negating the latency benefit of reuse.
The other challenge is \emph{cache maintenance}: existing methods use in-place cache updates assuming a static coordinate system; under per-block motion the cache drifts out of alignment, and writing recomputed values back along MVs creates write conflicts (e.g., multiple blocks are mapped to a same cache position) and staleness (e.g., cache positions that no blocks are mapped to are not updated). These errors compound across frames, progressively eroding reuse opportunities until caching offers no benefit over dense execution.

We propose FluxShard, a motion-aware edge-cloud video analytics system that achieves low latency and high accuracy by managing feature cache reuse at the granularity of individual motion regions.
FluxShard allows cached features to flow along their per-block motion trajectories to the positions where they best match the current frame; only positions whose post-alignment difference still exceeds a per-layer tolerance need recomputation.
For receptive field inconsistency, FluxShard establishes geometric conditions on the MV field under which cached convolution outputs remain valid despite heterogeneous motion, replacing the expensive per-layer feature comparison with a single lightweight pass over the input-level MV field. This is referred to as the Receptive Field Alignment Principle (RFAP).
For cache maintenance, FluxShard borrows the backward-MV warp from video codec reference frame reconstruction and applies it at the feature level: the entire cache is realigned with the current coordinate system and merged with freshly computed values, sustaining a high reuse ratio across consecutive frames.
The per-layer tolerances for cache reuse are jointly calibrated offline to find the largest values that minimize the recomputation workload while keeping accuracy within a prescribed budget.
A profiling-driven dispatcher then estimates per-endpoint latency from the resulting workload size and current network conditions, routing each frame to the lower-latency endpoint.

In summary, this paper makes the following contributions:
\begin{itemize}
	\item We identify the granularity mismatch between whole-scene cache management and local heterogeneous motion as the key limitation of existing feature reuse methods, and show that codec-level motion vectors provide a free, per-block geometric signal to overcome it.
	\item We propose FluxShard, a motion-aware edge-cloud video analytics system that manages feature cache reuse at the granularity of individual motion regions and routes the resulting recomputation workload to the lower-latency endpoint.
	\item We develop the RFAP to guarantee reuse correctness, motion-aware cache remapping to maintain cache validity across frames, and a profiling-driven dispatch scheduler to minimize per-frame latency.
	\item Evaluation on real-world dynamic video benchmarks shows that FluxShard achieves 32.6--83.8\% latency reduction and 14.9--64.0\% energy savings over all baselines while retaining  97.3--98.0\% of dense-model accuracy.
\end{itemize}

In the rest of this paper,
\S\ref{sec:motivation} motivates the problem of interest with empirical evidence.
\S\ref{sec:overview} defines the system model and pipeline.
\S\ref{sec:design} develops the reuse criterion, RFAP, cache remapping, and dispatch mechanisms.
\S\ref{sec:eval} evaluates FluxShard against four baselines.
\S\ref{sec:related_work} discusses the related work and \S\ref{sec:conclusion} concludes.
\section{Challenges and Motivation}\label{sec:motivation}

\subsection{The Edge-Cloud Dilemma}\label{sec:mot:dilemma}

Edge-cloud video analytics faces two competing bottlenecks: local compute and network transmission.
For example, running YOLO11m-pose~\cite{yolov11} on an NVIDIA Jetson Xavier NX requires approximately 446.8\,ms per $1024\times1024$ frame, far exceeding real-time requirements.
Offloading every frame eliminates the compute bottleneck but introduces a transmission bottleneck: an uncompressed $1024\times1024$ frame is about 3\,MB; at 30\,fps this sustains $\sim$755\,Mbps, routinely exceeding edge uplink capacity.
The cloud server completes inference within only about 27.6\,ms, yet end-to-end offload latency reaches approximately 290.3\,ms under a typical 5G uplink~\cite{narayanan_variegated_2021} ($\sim$383\,Mbps).
Figure~\ref{fig:mot:dilemma} quantifies this tension.

Feature cache reuse can reduce both bottlenecks: when consecutive frames share content, cached results substitute for fresh transmission and computation.
Existing mechanisms~\cite{spinn, coach, deltacnn,motiondeltacnn} treat caches in a fixed, whole-scene coordinate system, whether the reuse decision is binary (reuse or recompute the entire frame) or pixel-level (propagate a dense difference map).
This granularity mismatch breaks down under motion.
Figure~\ref{fig:mot:reuse} shows that even modest displacement triggers widespread invalidation: the reuse ratio of DeltaCNN~\cite{deltacnn} and M-DeltaCNN~\cite{motiondeltacnn} drops from over 60\% on near-static sequences to below 25\% once motion exceeds 20\,px/frame, despite most content being merely shifted rather than changed.

\noindent\textbf{Takeaway.}
Cache reuse is essential, but whole-scene granularity cannot tolerate motion. A finer-grained, motion-aware mechanism is needed.

\subsection{Motion Vectors: Opportunity and Correctness Failure}\label{sec:mot:mv}

\begin{figure}[t]
	\centering
	\subfloat[Edge-cloud dilemma]{\includegraphics[width=0.48\columnwidth]{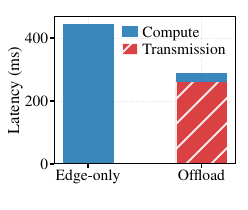}\label{fig:mot:dilemma}}%
	\hfill
	\subfloat[Reuse ratio vs.\ motion]{\includegraphics[width=0.48\columnwidth]{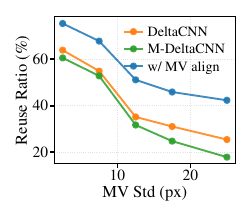}\label{fig:mot:reuse}}\\[2pt]
	\subfloat[Accuracy under naive reuse]{\includegraphics[width=0.48\columnwidth]{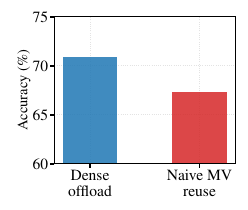}\label{fig:mot:accuracy}}%
	\hfill
	\subfloat[Naive cache update drift]{\includegraphics[width=0.48\columnwidth]{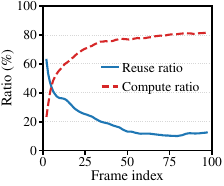}\label{fig:mot:bleed}}%
	\caption{Motivation for FluxShard.
	All panels use pose estimation (YOLO11m-pose) on 3DPW under medium bandwidth with an NVIDIA Jetson Xavier NX as the edge device and an RTX 3080 as the cloud server.
	(a)~End-to-end latency for dense local vs.\ dense offload inference.
	(b)~Reuse ratio vs.\ per-frame motion intensity (MV standard deviation) for DeltaCNN, M-DeltaCNN, and MV-aligned reuse.
	(c)~Accuracy of na\"ive MV reuse without correctness checking vs.\ dense offload.
	(d)~Reuse ratio and compute ratio over the first 100 frames of a representative sequence without cache remapping.}
	\label{fig:mot:challenge}
\end{figure}

H.264/H.265 codecs estimate a per-block displacement as part of normal encoding, providing per-region motion information at no additional cost. Although lossy codec stages (transform, quantization) can distort feature statistics~\cite{galteri_deep_2017,duan_video_2020,choi_scalable_2022}, the motion vectors are a purely geometric signal unaffected by these stages. By shifting cached feature blocks according to these MVs, displaced content is realigned before differencing, and the remaining recomputation set concentrates on regions that MVs cannot explain: dis-occlusion, deformation, and newly appearing content.
As shown in Figure~\ref{fig:mot:reuse}, MV alignment recovers much of the lost reuse, maintaining over 55\% even under strong motion where DeltaCNN and M-DeltaCNN fall below 25\%.
These results establish that MV alignment can recover most of the displaced content; the remaining question is whether the aligned features are \emph{correct} for downstream inference.

Despite the improved reuse ratio, directly reusing MV-aligned features still produces severe accuracy loss.
Figure~\ref{fig:mot:accuracy} shows that accuracy (measured by Object Keypoint Similarity, OKS) drops from 70.95\% to 67.31\% under na\"ive MV reuse, a 5.13\% relative degradation.
The root cause is receptive field inconsistency: layers whose receptive field spans more than one position aggregate spatial neighborhoods.
When adjacent blocks carry different MVs, re-indexing assembles patches that were never contiguous in the original frame; the cached output was computed from a different neighborhood.
The resulting error compounds through cascaded layers.

\noindent\textbf{Takeaway.}
MV alignment eliminates most false cache misses, but exploiting it requires a correctness guarantee under heterogeneous per-block motion.
This is the first challenge our design needs to address.

\subsection{Na\"ive Cache Update Breaks Future Reuse}\label{sec:mot:sparsity}

Even with a correct recomputation set, the system still faces a second problem: how
to maintain the cache after sparse recomputation.
Existing reuse systems simply overwrite the cache in place.
This policy assumes that the cache and the current frame still share the same
coordinate system.
Under heterogeneous per-block motion, however, the cache remains anchored to
the previous frame while the current content has moved to new locations.

A straightforward extension is to write each newly recomputed
recomputed block back to the cache location indicated by its MV.
This also fails: the mapping from current blocks to cached
locations is many-to-one and incomplete, creating write conflicts at
overlapping targets and leaving stale content at unmapped locations.
Positions that should remain reusable cannot find a valid
history entry, while positions declared reusable may read an
incorrect value.

Once written into the history, these errors accumulate.
Figure~\ref{fig:mot:bleed} tracks this on a representative 1660-frame pose estimation sequence.
Without proper cache realignment, the reuse ratio drops from over 70\% to below 15\% and the compute ratio rises from 13\% to over 80\%, approaching dense execution cost.
Genuinely reusable content persists but no longer aligns with the current coordinate system.

\noindent\textbf{Takeaway.}
Correct recomputation set extraction is not enough.
MV alignment can isolate a small recomputation set. 
However, without first realigning the
cache to the current frame layout, sparse updates leave behind stale or
conflicting history.
The broken history enlarges the next recomputation set and corrupts nominal reuse.
Cache maintenance under heterogeneous motion is therefore the second
challenge that our design needs to address.

\section{System Overview}\label{sec:overview}
This section first defines the system model and optimization objective (\S\ref{sec:model}), and then describes the end-to-end pipeline (\S\ref{sec:pipeline}).

\subsection{System Model}\label{sec:model}

We consider an edge-cloud video analytics system in which an edge device (e.g., an NVIDIA Jetson) receives a live video stream $\{I_t\}_{t=1}^{T}$ from a co-located camera, and a remote cloud server provides additional compute capacity.
The two endpoints communicate over a time-varying bandwidth-limited uplink.
A block-level MV field $\mathbf{m}_t$ is extracted from the video stream for each frame $I_t$.
For every pixel position $(i,j)$ in $I_t$, $\mathbf{m}_t(i,j)$ gives the displacement to its corresponding position $(\hat\imath,\hat\jmath) = (i,j) - \mathbf{m}_t(i,j)$ in the reference frame $I_{t-1}$; all pixels within the same $16\!\times\!16$ block share one displacement.
Each endpoint hosts an identical copy of the inference model and maintains a per-layer feature cache of intermediate features from its most recent inference.
For each incoming frame $I_t$, FluxShard executes the recomputation workload at one endpoint: either locally at the edge or remotely at the cloud.

\noindent\textbf{Inference and feature cache.} The DNN $\mathcal{N}$ comprises $L$ layers indexed by $l \in \{1,\dots,L\}$.
Layer~$l$ takes an input feature map $\mathbf{F}_l \in \mathbb{R}^{H_l' \times W_l' \times C_l'}$ and produces an output feature map $\mathbf{O}_l \in \mathbb{R}^{H_l \times W_l \times C_l}$; by convention $\mathbf{F}_1 = I_t$.
The output at spatial position $(i,j)$ depends on a set of input positions called its receptive field, denoted $\mathcal{R}^l(i,j) \subseteq \{1,\dots,H_l'\} \times \{1,\dots,W_l'\}$, with radius~$r_l$.
For convolutional and linear layers, $\mathbf{w}^l$ denotes the weight tensor applied over this receptive field, and $s^l$ denotes the spatial stride.
Because our analysis only concerns spatial positions, we omit the channel dimension hereafter.
A contiguous group of feature-map positions (e.g., $16\!\times\!16$ pixels at the input) that correspond to a single MV entry is called a \emph{shard}.

At layer~$l$, $\hat{\mathbf{F}}_l$ and $\hat{\mathbf{O}}_l$ denote the cached input and output features; since $\hat{\mathbf{O}}_l \equiv \hat{\mathbf{F}}_{l+1}$, we use whichever is more natural in context.
At deeper layers, FluxShard downsamples the MV field from $\hat{\mathbf{m}}_0$ to match each layer's spatial resolution.

At each layer, FluxShard forms an \emph{assembled output} $\tilde{\mathbf{O}}_l$ by combining MV-aligned cached values with freshly computed ones.
Positions whose output discrepancy exceeds a per-layer tolerance $\tau_l$ require fresh computation and form the \emph{recomputation set} $\mathcal{S}_l$.
The thresholds $\boldsymbol{\tau} = \{\tau_l\}_{l=0}^{L}$ are calibrated offline, where $l\!=\!0$ denotes a virtual layer that FluxShard prepends to the DNN for estimating the per-endpoint recomputation workload; the exact reuse criterion and calibration procedure are developed in the next section.

Because both endpoints execute the same inference logic, we present the reuse criterion and cache update rules without endpoint superscripts; $\mathcal{S}_l$, $\hat{\mathbf{F}}_l$, and $\hat{\mathbf{O}}_l$ refer to the selected endpoint's state.
Superscripts $e$ or $c$ appear only when the dispatch scheduler compares both endpoints.

\noindent\textbf{Optimization objective.}
Edge intelligence applications such as autonomous navigation, robotic manipulation, and augmented-reality interaction often require both real-time responsiveness and high perception accuracy; neither can be sacrificed for the other.
Let $A_t$ denote the task accuracy (e.g., mAP) achieved on frame $I_t$ under the realized reuse and endpoint selection; $A_t^*$ denotes the accuracy of full recomputation without caching; $T_t$ denotes the corresponding per-frame end-to-end latency, comprising local computation, data transmission, and remote computation as applicable.
We define the sequence-level averages: $\bar{A} = \frac{1}{T}\sum_{t=1}^{T} A_t$, $\bar{A}^* = \frac{1}{T}\sum_{t=1}^{T} A_t^*$, and $\bar{T} = \frac{1}{T}\sum_{t=1}^{T} T_t$.
This leads to the following optimization objective:
\begin{equation}\label{eq:objective}
	\min_{\{d_t\}} \; \bar{T}
	\quad \text{s.t.} \quad
	\bar{A} \;\ge\; \alpha \,\bar{A}^*,
\end{equation}
where $d_t \in \{edge, cloud\}$ is the per-frame dispatch decision and $\alpha \in (0,1]$ is a user-specified accuracy retention ratio.
The objective minimizes average latency over the video sequence while keeping the accuracy loss within a user-specified budget.


\subsection{Pipeline}\label{sec:pipeline}
\begin{figure}[t]
	\centering
	\includegraphics[width=\linewidth]{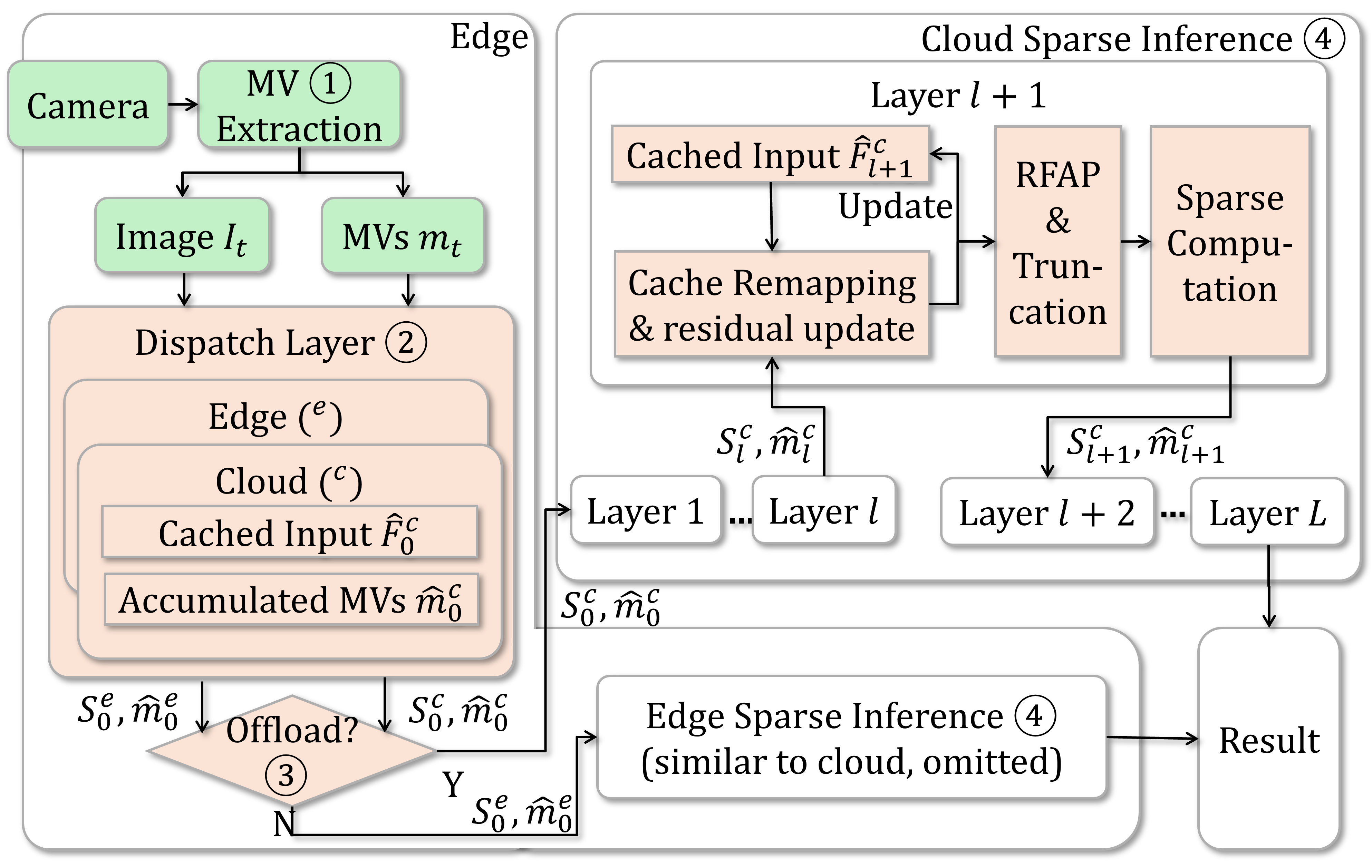}
	\caption{Overview of the FluxShard pipeline.}\label{fig:overview}
\end{figure}

Figure~\ref{fig:overview} illustrates the pipeline.
FluxShard introduces a dispatch layer (virtual layer~$0$) preceding the DNN.
This layer applies the identity operator to the raw input, so its reuse criterion reduces to a direct per-position comparison between the current frame and the MV-aligned cached input.
For both endpoints, it stores a cached input ($\hat{\mathbf{F}}_0^{e}$, $\hat{\mathbf{F}}_0^{c}$) and an accumulated MV field ($\hat{\mathbf{m}}_0^{e}$, $\hat{\mathbf{m}}_0^{c}$) that tracks the total displacement from the cached input to the present frame.
When an encoded frame arrives at the edge, FluxShard proceeds in the following four stages.

\emph{Stage 1: MV extraction.}
Block-level motion vectors $\mathbf{m}_t$ are extracted from the codec during decoding at negligible cost.
Both dispatch states update their accumulated MV fields by warping the old accumulator along the new MV and adding the new displacement.
When an endpoint performs inference, its accumulated field is reset; the other endpoint's field continues to accumulate.
At deeper DNN layers, the MV field is downsampled from $\hat{\mathbf{m}}_0$ to match each layer's spatial resolution; we write $\hat{\mathbf{m}}_l$ for the field on the grid of layer~$l$'s output (equivalently, layer~$l\!+\!1$'s input).

\emph{Stage 2: Recomputation workload estimation.}
Each dispatch state uses its accumulated MV field to shift the cached input shard by shard, aligning each shard to its current-frame position.
A per-position comparison within each aligned shard identifies positions whose content differs from the cache by more than the dispatch threshold $\tau_0$; these positions form the candidate input recomputation sets $\mathcal{S}_0^{e}$ and $\mathcal{S}_0^{c}$.
From their sizes, the edge state estimates local sparse inference latency, while the cloud state estimates the cost of transmitting the MV field and recomputation pixels plus remote inference under current network conditions.

\emph{Stage 3: Dispatch decision.}
The frame is assigned to the endpoint with the lower estimated latency; the accuracy constraint in Eq.~\eqref{eq:objective} is enforced by the profiled thresholds $\boldsymbol{\tau}$ rather than by the routing step itself.

\emph{Stage 4: Inference and cache update.}
After dispatch, only the selected endpoint continues sparse inference.
The selected endpoint begins from the layer-0 recomputation set and propagates it through the DNN one layer at a time.
At each layer, the system remaps the cached features to the current-frame coordinate system, identifies which positions can be safely reused and which must be recomputed, and sparsely evaluates only the latter.
Freshly computed and cached values are assembled into a result that serves as both the next layer's input and the updated cache.
After the final layer, the endpoint resets its accumulated MV field.

The non-selected endpoint retains its accumulated MV field for future reuse.
When the cloud is selected, the edge transmits only $\hat{\mathbf{m}}_0^{c}$ and the content-change recomputation pixels in $\mathcal{S}_0^{c}$; the cloud applies the RFAP check from the MV field before starting DNN inference.

These four stages rely on three mechanisms developed in the next section: per-layer thresholds $\boldsymbol{\tau}$ that control the accuracy-latency tradeoff by determining the recomputation set size, RFAP that ensures reuse correctness under heterogeneous motion, and cache remapping that keeps the reference aligned across frames.

\section{System Design}\label{sec:design}

This section describes how FluxShard realizes the optimization objective in Eq.~\eqref{eq:objective}.
We first reduce it to a per-frame recomputation minimization problem based on cache reuse (\S\ref{sec:residual-min}).
We then formalize the per-position reuse criterion (\S\ref{sec:criterion}) and present RFAP, which makes MV-aligned reuse correct under heterogeneous motion (\S\ref{sec:rfap}).
Next, we describe threshold calibration and cache remapping, which shrink the recomputation workload and preserve a valid cache reference respectively (\S\ref{sec:minimize}).
Finally, we integrate these mechanisms into a per-frame pipeline with dispatch (\S\ref{sec:dispatch}).

\subsection{From Design Goal to Minimal Feasible Recomputation Set}\label{sec:residual-min}

The optimization objective (Eq.~\eqref{eq:objective}) minimizes average latency $\bar{T}$, subject to $\bar{A}\ge\alpha\bar{A}^*$.
In principle, the optimal solution would couple all frames: each frame's reuse and dispatch decisions determine the cache state inherited by subsequent frames, so minimizing $\bar{T}$ globally requires knowledge of future motion and content.
Since in an online streaming setting this knowledge is unavailable, FluxShard relaxes Eq.~\eqref{eq:objective} into a per-frame greedy formulation.
Requiring $A_t \ge \alpha\,A_t^{*}$ at every frame is a sufficient condition for the sequence-level accuracy constraint, and minimizing $T_t$ per frame yields:
\begin{equation}\label{eq:per-frame}
\min_{\mathcal{S}^e,\,\mathcal{S}^c} \; \min\!\bigl\{T_t^e(\mathcal{S}^e),\; T_t^c(\mathcal{S}^c,\, \hat{B})\bigr\}
\quad\text{s.t.}\quad A_t \;\ge\; \alpha\, A_t^{*},
\end{equation}
where $\mathcal{S}^e = \{\mathcal{S}_l^e\}_{l=0}^{L}$ and $\mathcal{S}^c = \{\mathcal{S}_l^c\}_{l=0}^{L}$ are the collections of per-layer recomputation sets on each endpoint, computed over independent cache states, and $\hat{B}$ is the current uplink bandwidth estimate.
The inner $\min$ is the dispatch decision: the frame is routed to the faster endpoint.
The outer minimization requires each endpoint to minimize its own $T_t$.
Since per-endpoint latency is monotonically non-decreasing in the recomputation set size (i.e., the edge path incurs sparse computation determined by $\{\mathcal{S}_l^e\}$, while the cloud path additionally incurs transmission of $\mathcal{S}_0^c$), minimizing $T_t^e$ and $T_t^c$ each reduces to minimizing the recomputation sets:
\begin{equation}\label{eq:design-subproblem}
\min_{\mathcal{S}_0,\,\{\mathcal{S}_l\}} \; |\mathcal{S}_0| + \textstyle\sum_{l} |\mathcal{S}_l|
\quad\text{s.t.}\quad A_t \;\ge\; \alpha\, A_t^{*}.
\end{equation}
Because both endpoints execute the same reuse logic, Eq.~\eqref{eq:design-subproblem} applies to each independently; the system constructs the smallest feasible recomputation workload on each endpoint and routes the frame to the lower-latency one.

\subsection{Reuse Criterion}\label{sec:criterion}

The core of Eq.~\eqref{eq:design-subproblem} is deciding, at each layer and each position, whether the cached value can substitute for fresh computation without violating the accuracy constraint.
As shown in \S\ref{sec:mot:mv}, MV alignment removes displacement-induced false changes and improves cache reuse, but na\"ive reuse of MV-aligned features still degrades accuracy.
This subsection establishes a per-position reuse criterion that formalizes when reuse is safe and identifies the source of the accuracy loss.

As introduced in \S\ref{sec:model}, three quantities coexist at every output position $(i,j)$ of layer $l$:
the ground-truth output $\mathbf{O}_l(i,j)$, obtained by applying layer $l$ to the current input $\mathbf{F}_l$ (or, in practice, to the assembled input $\tilde{\mathbf{O}}_{l-1}$);
the cached output $\hat{\mathbf{O}}_l(\hat{\imath},\hat{\jmath})$ from a previous frame, computed from $\mathcal{R}^l(\hat{\imath},\hat{\jmath})$ in the cached input;
and the assembled output $\tilde{\mathbf{O}}_l(i,j)$ is consumed by layer $l\!+\!1$.
Reuse at position $(i,j)$ is \emph{valid} when substituting the cached value introduces bounded error:
\begin{equation}\label{eq:reuse-output}
	\bigl\lvert\,\mathbf{O}_{l}(i,j)
	- \hat{\mathbf{O}}_{l}(\hat{\imath},\hat{\jmath})
	\bigr\rvert
	\;\le\;\tau_{l},
\end{equation}
where setting $\tau_{l}=0$ reduces the criterion to exact output equality.
Based on this criterion, FluxShard assembles the output of layer $l$ as
\begin{equation}\label{eq:assemble}
	\tilde{\mathbf{O}}_l(i,j) =
	\begin{cases}
		\hat{\mathbf{O}}_l(\hat{\imath},\hat{\jmath}),
		 & \text{if Eq.~\eqref{eq:reuse-output} holds;} \\[3pt]
		\mathbf{O}_l(i,j),
		 & \text{otherwise.}
	\end{cases}
\end{equation}
The assembled output serves as the input to layer $l\!+\!1$.
Positions assigned to the ``otherwise'' branch form the recomputation set $\mathcal{S}_l$.

Eq.~\eqref{eq:reuse-output} cannot be checked directly without computing $\mathbf{O}_{l}(i,j)$, which defeats the purpose of reuse.
Because the cached output at $(\hat{\imath},\hat{\jmath})$ was produced from $\mathcal{R}^l(\hat{\imath},\hat{\jmath})$ in the cached input, reuse is safe whenever the current input within $\mathcal{R}^l(i,j)$ matches that cached patch.
For each $(p,q) \in \mathcal{R}^l(i,j)$, let $(\hat{p},\hat{q})$ denote the corresponding position in $\mathcal{R}^l(\hat{\imath},\hat{\jmath})$ (i.e., at the same kernel offset). We denote the maximum absolute input perturbation over all spatial positions in the receptive field at output position $(i,j)$ as
\begin{equation}\label{eq:delta-max}
	\Delta_{\max}^{l}(i,j)
	\;=\;
	\max_{(p,q)\,\in\,\mathcal{R}^l(i,j)}
	\bigl\lvert\,\mathbf{F}_{l}(p,q)
	- \hat{\mathbf{F}}_{l}(\hat{p},\hat{q})
	\bigr\rvert.
\end{equation}
For a linear layer with weight vector $\mathbf{w}^l$ (the kernel flattened over all $N$ input elements in the receptive field), linearity yields
\begin{equation}\label{eq:linear-bound}
	\bigl\lvert\,\mathbf{O}_{l}(i,j)
	- \hat{\mathbf{O}}_{l}(\hat{\imath},\hat{\jmath})
	\bigr\rvert
	\;\le\;
	\lVert\mathbf{w}^l\rVert_1 \cdot
	\Delta_{\max}^{l}(i,j),
\end{equation}
so Eq.~\eqref{eq:reuse-output} is guaranteed whenever the input patch satisfies
\begin{equation}\label{eq:reuse-input}
	\Delta_{\max}^{l}(i,j)
	\;\le\;
	\frac{\tau_{l}}{\lVert\mathbf{w}^l\rVert_1}.
\end{equation}
For element-wise activations with Lipschitz constant at most~$1$ (ReLU, sigmoid, etc.), the input-side bound directly implies the output-side one.
Batch normalization layers are affine at inference time, so Eq.~\eqref{eq:linear-bound} applies directly.
Each layer's $\tau_{l}$ bounds the single-layer error independently; the cumulative effect of cascaded approximations is controlled by the offline profiling that jointly calibrates all $\tau_{l}$ over complete forward passes (discussed in \S\ref{sec:minimize}).

Evaluating Eq.~\eqref{eq:reuse-input} over the full receptive field at every output position and every layer costs as much as executing the layer itself.
Per-position reuse from the previous layer can reduce this cost: for $(p,q) \notin \mathcal{S}_{l-1}$, the assembled value equals the cached value at the position indicated by the input MV $\hat{\mathbf{m}}_{l-1}(p,q)$, contributing zero to $\Delta_{\max}^l$.
We call \emph{reuse propagation} the practice of treating a position as reusable whenever its entire receptive field falls outside $\mathcal{S}_{l-1}$, which confines the evaluation of Eq.~\eqref{eq:reuse-input} to the neighborhood of recomputation positions.
For layers with receptive field size one (pointwise convolutions, element-wise activations), this per-position check is sufficient: each output depends on exactly one input, so reuse at a position in the previous layer directly carries over to the same position at the current layer.
For layers with receptive field size greater than one (e.g., $3\!\times\!3$ convolutions), however, output reuse at $(i,j)$ requires the current receptive field $\mathcal{R}^l(i,j)$ to match the cached receptive field $\mathcal{R}^l(\hat{\imath},\hat{\jmath})$, and zero per-position input difference can only imply zero receptive field level difference if the input MV is uniform within the receptive field and geometrically coherent with the output MV.

\subsection{Receptive Field Alignment Principle}\label{sec:rfap}

Under heterogeneous per-block MVs, the above MV conditions are frequently violated, preventing reuse from propagating across layers with receptive field size greater than one.
As shown in \S\ref{sec:mot:mv}, ignoring such violations causes significant accuracy loss.
To restore safe reuse propagation, the system must identify positions that violate the MV conditions and force them into the recomputation set.
RFAP provides two sufficient conditions on the input-level MV field $\hat{\mathbf{m}}_0$ that identify all such positions in a single pass.

The first condition, \emph{intra-receptive-field uniformity}, requires all input positions within the receptive field to share a single displacement:
\begin{equation}\label{eq:cond-uniform}
	\hat{\mathbf{m}}_{l-1}(p,q)
	= \hat{\mathbf{m}}_{l-1}(p',q'),
	\quad
	\forall\,(p,q),\,(p',q') \in \mathcal{R}^l(i,j).
\end{equation}
A uniform shift ensures that the per-position cached values form the same contiguous patch $\mathcal{R}^l(\hat{\imath},\hat{\jmath})$ that produced the cached output.

The second condition, \emph{input-output geometric coherence}, requires 
\begin{equation}\label{eq:cond-geometric}
	\hat{\mathbf{m}}_{l} = \hat{\mathbf{m}}_{l-1} / s^l,
\end{equation}
where $s^l$ is the stride of layer $l$ (assuming uniform stride in both dimensions).
This ensures that the output MV and the (uniform) input MV point to the same cached content.
When the displacement value is not divisible by the stride, the two MV fields point to different cached positions, and the output must be recomputed.

When both conditions hold at a given output position, per-position reuse from the previous layer directly implies receptive field level correctness, and no further check is needed.
Positions that violate either condition are forced into the recomputation set for recomputation, preserving accuracy.

We then compact the evaluation of both conditions to the input level.
Both conditions can be evaluated from the input-level MV field $\hat{\mathbf{m}}_0$ alone, because deeper-layer fields are obtained by downsampling $\hat{\mathbf{m}}_0$.
For Condition~1, checking MV uniformity within the largest effective receptive field $R_{\max}$ (measured at the input grid) covers all layers: uniformity within $R_{\max}$ positions implies uniformity within any smaller receptive field at any intermediate layer.
For Condition~2, checking that the displacement is divisible by the largest cumulative stride $S_{\max} = \max_l \prod_{k=1}^{l} s^k$ covers all individual strides.
At the shard level, Condition~1 violations arise near shard boundaries where the $R_{\max}$-neighborhood spans multiple shards with different displacements; Condition~2 violations occur at shards whose displacement is indivisible by $S_{\max}$.

Input-level positions that fail either condition are merged into the recomputation set at the first DNN layer whose receptive field exceeds one.
Because the dispatch layer is an identity operator with receptive field size one, these positions do not affect $\mathcal{S}_0$.
At downstream layers, no separate MV check is needed: flagged positions are recomputed, and their fresh values propagate through the network via Eq.~\eqref{eq:reuse-input}.

With $\tau_l = 0$ at all layers, the compacted check is provably correct: every structurally inconsistent position is recomputed.
With profiled $\tau_l > 0$, some deeper-layer positions affected by resolved inconsistencies may exhibit small activation differences that truncation absorbs, avoiding unnecessary recomputation while staying within the calibrated accuracy budget.

\subsection{Minimizing the Recomputation Set}\label{sec:minimize}

The reuse criterion and RFAP establish how to efficiently determine reusable positions given a per-layer output discrepancy bound $\tau_l$ under heterogeneous motion.
However, the relationship between $\tau_l$ and task accuracy $A_t$ has not yet been established: larger thresholds yield smaller recomputation sets but risk violating the accuracy constraint in Eq.~\eqref{eq:design-subproblem}.
This subsection addresses this gap: threshold calibration finds the largest feasible $\tau_l$ that keeps accuracy within budget, thereby minimizing the recomputation set; cache remapping then keeps the cache aligned so that subsequent frames also start from a valid reference cache.

\subsubsection{Profiling-Driven Truncation}\label{sec:truncation}

Even with RFAP and MV alignment, exact reuse ($\tau_l = 0$) leaves many low-magnitude discrepancies with negligible task impact.
FluxShard assigns a nonzero $\tau_l$ to each profiled layer, truncating these discrepancies and reusing the cached value instead.
A recomputation position is truncated whenever it satisfies the same input-side bound as Eq.~\eqref{eq:reuse-input} with the calibrated $\tau_l$.
In practice, nonzero thresholds are profiled only for the dispatch layer and a subset of DNN layers $\mathcal{L}_{\mathrm{tr}}$ (selected activation layers); all other layers use $\tau_{l}=0$.
At layers with $\tau_l = 0$, only positions whose MV-aligned input has actually changed enter the recomputation set; the threshold merely forgoes the additional truncation of small changes, so sparsity remains high whenever the MV alignment successfully recovers displaced content.
At the dispatch layer, the same rule uses the identity operator, i.e., $\lVert\mathbf{w}\rVert_1 = 1$, to form the input recomputation set $\mathcal{S}_0$.

The thresholds are jointly calibrated offline on the training set of each task.
Let $\mathcal{L}_{\mathrm{tr}} \subseteq \{1,\dots,L\}$ denote the set of profiled DNN layers, and $K = |\mathcal{L}_{\mathrm{tr}}|$.
The total admissible accuracy drop relative to the dense baseline is
\begin{equation}\label{eq:truncation-budget}
	\Delta A = (1-\alpha) \bar{A}^{*}.
\end{equation}
Because the dispatch layer determines the input recomputation set and therefore all downstream workload---including the recomputation pixels transmitted when inference is offloaded---it has the greatest impact on latency and receives the larger share.
The 2:1 ratio is empirically tuned on the training set.
In other words, we reserve $\frac{2}{3}$ of the budget for $\tau_{0}$ and split the remaining $\frac{1}{3}$ evenly across the profiled DNN layers:
\begin{equation}\label{eq:truncation-budget-split}
	b_0 = \frac{2}{3}\Delta A,
	\qquad
	b_l = \frac{1}{3K}\Delta A,
	\;\; l \in \mathcal{L}_{\mathrm{tr}}.
\end{equation}
The dispatch layer is first processed, followed by the profiled DNN layers in the network order.
For each profiled stage $u \in \{0\} \cup \mathcal{L}_{\mathrm{tr}}$, we search a discrete candidate set $\mathcal{C}^{u}$ and choose the largest threshold whose cumulative accuracy drop remains within the cumulative budget assigned up to that stage.
This greedy search returns one calibrated $\tau_{0}$ for input recomputation extraction and one $\tau_{l}$ for each profiled DNN layer.
At runtime, the dispatch layer applies $\tau_{0}$ to form $\mathcal{S}_0$, and each profiled DNN layer applies its calibrated $\tau_{l}$ using the same criterion.

\subsubsection{Motion-Aware Cache Remapping}\label{sec:cache_remap}

Threshold calibration minimizes the current frame's recomputation set, but subsequent frames depend on how well the cache reflects the current frame.
Without realignment after each inference, the cache becomes stale: reusable positions no longer align with valid history, inflating the recomputation set, and positions declared reusable may consume corrupted cached features.
As shown in \S\ref{sec:mot:sparsity}, this drift causes the recomputation set to inflate over time, eventually exceeding the cost of dense recomputation.

In video codecs, a reference frame is reconstructed by warping the previous reference along backward MVs and then merging newly decoded residuals; each position maps to exactly one source, so the warp is conflict-free.
FluxShard applies this principle at the feature level: each current-frame shard looks up its source region in the cached features via backward MVs, then merges freshly computed values.
At layer $l\!+\!1$, the cache update proceeds in two steps.
First, the cached input $\hat{\mathbf{F}}_{l+1}$ is warped to the current-frame coordinate system using the accumulated MV field:
\begin{equation}\label{eq:cache-warp}
	\hat{\mathbf{F}}_{l+1,\mathrm{remap}}(i,j)
	\;=\;
	\hat{\mathbf{F}}_{l+1}\!\bigl(
	(i,j) - \hat{\mathbf{m}}_{l}(i,j)
	\bigr).
\end{equation}
Because the backward MV assigns each position to exactly one source, the warp is conflict-free with no unwritten holes.
Second, the remapped cache is merged with the freshly computed outputs from layer $l$:
\begin{equation}\label{eq:cache-merge}
	\hat{\mathbf{F}}_{l+1}(i,j)
	\;=\;
	\begin{cases}
		\mathbf{O}_l(i,j),
		 & (i,j) \in \mathcal{S}_l, \\[3pt]
		\hat{\mathbf{F}}_{l+1,\mathrm{remap}}(i,j),
		 & \text{otherwise.}
	\end{cases}
\end{equation}
Recomputation positions receive freshly computed values; reusable positions retain the warped cache.
The resulting tensor replaces $\hat{\mathbf{F}}_{l+1}$ (equivalently $\hat{\mathbf{O}}_l$) as the new cached state.

After the update, the cache resides in the current-frame coordinate system, and the endpoint resets its accumulated MV field.
Subsequent lookups start from identity alignment, avoiding double-shifting the remapped cache.
The same remap-then-update rule applies to the input cache and to all intermediate layer caches maintained by the sparse runtime.

\subsection{Per-Frame Pipeline and Dispatch}\label{sec:dispatch}

\SetAlgoNlRelativeSize{-1}
\begin{algorithm}[t]
	\caption{FluxShard Per-Frame Pipeline}\label{alg:pipeline}
	\KwIn{$I_t$;
	$(\hat{\mathbf{F}}_0^{e},\,\hat{\mathbf{m}}_0^{e})$;
	$(\hat{\mathbf{F}}_0^{c},\,\hat{\mathbf{m}}_0^{c})$;
	$\tau_0$, $\boldsymbol{\tau}$;
	$f_{\mathrm{edge}}$,
	$f_{\mathrm{cloud}}$}
	\KwOut{Inference result $\mathbf{y}_t$}
	\tcp{Stage 1: MV extraction}
	Extract MV field $\mathbf{m}_t$ from codec for frame
	$I_t$\;
	$\hat{\mathbf{m}}_0^{e} \gets
		\texttt{Accumulate}(\hat{\mathbf{m}}_0^{e},\,\mathbf{m}_t)$\;
	$\hat{\mathbf{m}}_0^{c} \gets
		\texttt{Accumulate}(\hat{\mathbf{m}}_0^{c},\,\mathbf{m}_t)$\;
	\tcp{Stage 2: Recomputation workload estimation}
	$\mathcal{S}_0^{e} \gets \texttt{RecomputeSet}(I_t,\,\hat{\mathbf{F}}_0^{e},\,\hat{\mathbf{m}}_0^{e},\,\tau_0)$
		\tcp*{Eq.~\eqref{eq:residual-set-dispatch}}
	$\mathcal{S}_0^{c} \gets \texttt{RecomputeSet}(I_t,\,\hat{\mathbf{F}}_0^{c},\,\hat{\mathbf{m}}_0^{c},\,\tau_0)$\;
	$\rho_e \gets |\mathcal{S}_0^{e}| / |I_t|$;\quad
	$\rho_c \gets |\mathcal{S}_0^{c}| / |I_t|$\;
	$T_{\mathrm{edge}} \gets
		f_{\mathrm{edge}}(\rho_e)$;\quad
	$T_{\mathrm{cloud}} \gets
		f_{\mathrm{cloud}}(\rho_c)
		+ |\mathcal{S}_0^{c}|\,/\,\hat{B}$\;
	\tcp{Stage 3: Dispatch decision}
	\eIf{$T_{\mathrm{edge}} < T_{\mathrm{cloud}} - \epsilon$}{
	\tcp{Stage 4: Inference and cache update (edge)}
	$\mathcal{S}_{\mathrm{rfap}} \gets \texttt{RFAP}(\hat{\mathbf{m}}_0^{e})$
		\tcp*{Eq.~\eqref{eq:cond-uniform}}
	$\mathbf{y}_t \gets
		\texttt{SparseInfer}_{\mathrm{edge}}(
		\mathcal{S}_0^{e} \cup \mathcal{S}_{\mathrm{rfap}},\,\boldsymbol{\tau})$\;
	Update edge cache via Eqs.~\eqref{eq:cache-warp}--\eqref{eq:cache-merge};\quad
	reset $\hat{\mathbf{m}}_0^{e}$\;
	}{
	\tcp{Stage 4: Inference and cache update (cloud)}
	Transmit $\mathcal{S}_0^{c}$ and $\hat{\mathbf{m}}_0^{c}$ to
	cloud\;
	$\mathcal{S}_{\mathrm{rfap}} \gets \texttt{RFAP}(\hat{\mathbf{m}}_0^{c})$\;
	$\mathbf{y}_t \gets
		\texttt{SparseInfer}_{\mathrm{cloud}}(
		\mathcal{S}_0^{c} \cup \mathcal{S}_{\mathrm{rfap}},\,\boldsymbol{\tau})$\;
	Update cloud cache via Eqs.~\eqref{eq:cache-warp}--\eqref{eq:cache-merge};\quad
	reset $\hat{\mathbf{m}}_0^{c}$\;
	}
	\Return $\mathbf{y}_t$\;
\end{algorithm}

The preceding subsections solve Eq.~\eqref{eq:design-subproblem} on each endpoint: the reuse criterion and RFAP enforce correctness, threshold calibration minimizes the recomputation sets under the accuracy constraint, and cache remapping maintains validity across frames.
The remaining step in Eq.~\eqref{eq:per-frame} is the dispatch: selecting the endpoint with lower $T_t$.

For each incoming frame, both dispatch-layer states (\S\ref{sec:pipeline}) incorporate the new MV field $\mathbf{m}_t$ into their accumulated fields:
\begin{equation}\label{eq:mv-accumulate}
	\hat{\mathbf{m}}_0(i,j)
	\;=\;
	\hat{\mathbf{m}}_0\!\bigl((i,j) - \mathbf{m}_t(i,j)\bigr)
	\;+\; \mathbf{m}_t(i,j),
\end{equation}
where the composition warps the old accumulator to the current coordinate system before adding the new displacement.
Each state then extracts its recomputation set by applying the reuse criterion with the dispatch-layer threshold $\tau_{0}$ and the identity operator ($\lVert\mathbf{w}\rVert_1 = 1$):
\begin{equation}\label{eq:residual-set-dispatch}
	\mathcal{S}_0
	\;=\;
	\bigl\{(i,j) :
	\lvert\, I_t(i,j)
	- \hat{\mathbf{F}}_0\!\bigl((i,j) - \hat{\mathbf{m}}_0(i,j)\bigr)
	\rvert > \tau_{0}
	\bigr\}.
\end{equation}

Let $\rho_e = |\mathcal{S}_0^{e}|/N_{\mathrm{px}}$ and $\rho_c = |\mathcal{S}_0^{c}|/N_{\mathrm{px}}$ denote the normalized recomputation set sizes.
The bandwidth estimate $\hat{B}$ is maintained as an exponentially weighted moving average of recent uplink measurements.
FluxShard profiles the latency-vs-sparsity relationship offline on each endpoint, yielding:
\begin{align}
	T_{\mathrm{edge}} &\;=\; f_{\mathrm{edge}}(\rho_e),
	\label{eq:latency-edge} \\
	T_{\mathrm{cloud}} &\;=\; f_{\mathrm{cloud}}(\rho_c)
	\;+\; |\mathcal{S}_0^{c}| \,/\, \hat{B},
	\label{eq:latency-cloud}
\end{align}
where $f_{\mathrm{edge}}$ and $f_{\mathrm{cloud}}$ are the profiled curves.
The scheduler selects the endpoint with lower total latency; when the two estimates are within a margin $\epsilon$, it prefers cloud offload to reduce edge energy.

After dispatch, only the selected endpoint runs inference.
The RFAP check is applied on the selected endpoint's MV field to identify structurally inconsistent positions, which are merged with $\mathcal{S}_0$ before sparse inference begins.
Because the dispatch layer has receptive field size one, RFAP-flagged positions are pixel-level correct after MV alignment and need not be transmitted; when inference is offloaded, only $\mathcal{S}_0^{c}$ and $\hat{\mathbf{m}}_0^{c}$ are sent to the cloud, and the cloud applies the RFAP check locally from the received MV field.
After inference, the cache is updated via Eqs.~\eqref{eq:cache-warp}--\eqref{eq:cache-merge} and the accumulated MV field is reset.

Algorithm~\ref{alg:pipeline} summarizes the complete per-frame pipeline, integrating MV extraction, workload estimation, dispatch, inference, and cache update into the four stages described in \S\ref{sec:pipeline}.
Lines~1--3 extract and accumulate the MV field into both endpoint states.
Lines~4--7 compute the content-change recomputation sets and estimate per-endpoint latency.
Lines~8--16 dispatch to the cheaper endpoint, apply the RFAP check (line~9, 14), run sparse inference with profiled thresholds $\boldsymbol{\tau}$ (line~10, 15), and update the cache (line~11, 16).

\section{Evaluation}\label{sec:eval}

\subsection{Experimental Setup}\label{sec:eval-setup}

\begin{figure}[t]
    \centering
    \includegraphics[width=\columnwidth]{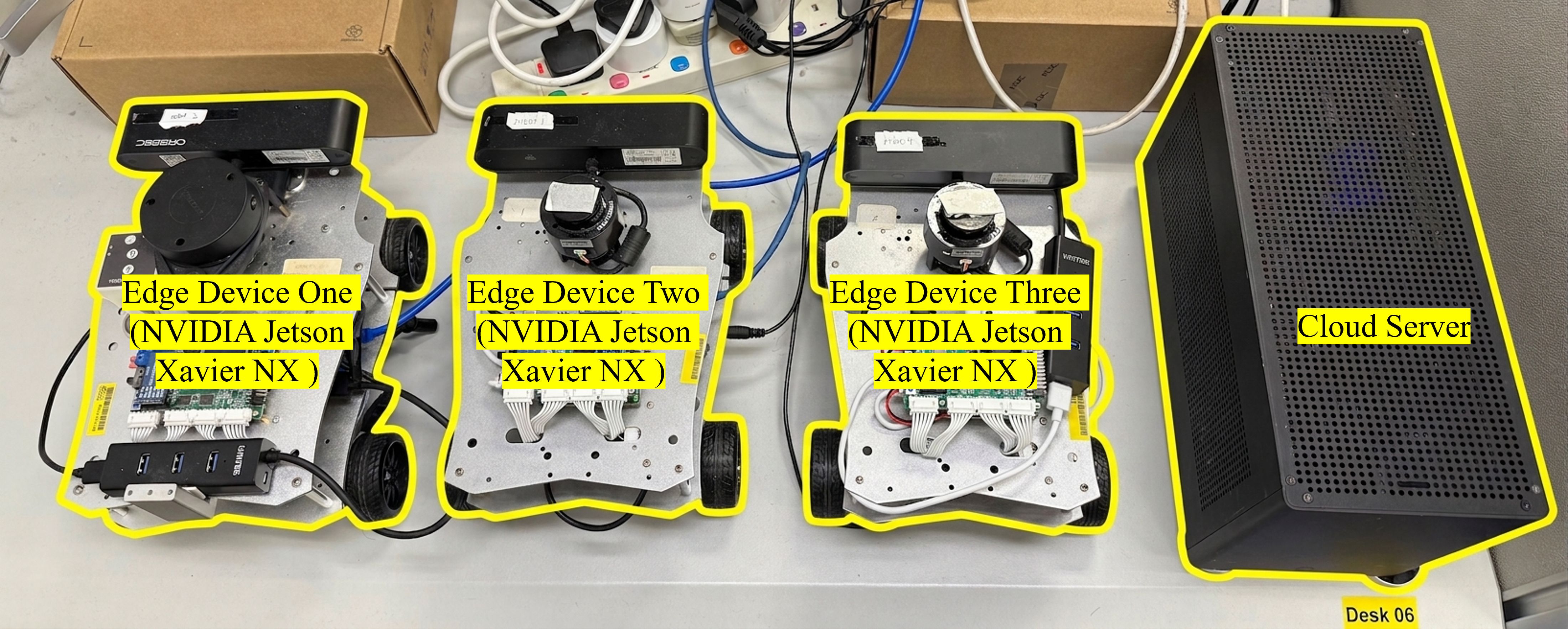}
    \caption{Testbed. Left: three mobile robots with embedded Jetson Xavier NX boards. Right: cloud server with RTX 3080.}
    \label{fig:testbed}
\end{figure}

\paragraph{Implementation and testbed}
FluxShard is implemented in Python and PyTorch, with custom CUDA kernels for sparse convolution, pooling, the RFAP consistency check, and activation with fused cache maintenance (truncation, MV-guided history lookup, and MV-guided cache remapping and updating in a single pass).
All feature tensors use channel-last (NHWC) layout to match the per-position sparse access pattern.
The edge client and cloud server communicate over a TCP socket; when a frame is offloaded, the client sends the accumulated MV field, a bitwise-packed and downsampled ($2\!\times\!2$) recomputation mask, and the recomputation RGB pixels.
The MV field ($16\times16$ block size) occupies approximately 0.52\% of the full RGB input, and the recomputation mask adds 1.04\%, totaling 1.56\% metadata overhead independent of input resolution.

Figure~\ref{fig:testbed} shows the testbed.
The edge devices are NVIDIA Jetson Xavier NX with 384 CUDA cores and 8\,GB LPDDR4x memory, embedded in mobile robots.
The cloud server hosts an NVIDIA RTX 3080 GPU with 10\,GB GDDR6X memory.
All machines run Ubuntu 20.04 with CUDA 11.4.
They are connected through a 1\,Gbps Ethernet switch, while the wireless uplink is emulated with Linux \texttt{tc}.

We replay client-to-server bandwidth traces from a public 4G/5G measurement dataset~\cite{narayanan_variegated_2021} and group them into three tiers: \emph{low}, \emph{medium}, and \emph{high}.
The \emph{low} tier uses 4G/LTE traces ($40.4\pm36.6$\,Mbps); the \emph{medium} and \emph{high} tiers are obtained by splitting the 5G traces at the median throughput into a lower half ($382.8\pm419.1$\,Mbps) and an upper half ($596.9\pm467.9$\,Mbps).
The emulated link also adds a fixed 20\,ms one-way propagation delay.
Unless otherwise noted, we report results under the medium tier.

\paragraph{Workloads and datasets}
We evaluate two real-time dense prediction workloads with the YOLO11 family~\cite{yolov11}.
We choose YOLO11 as a representative evaluation model because its backbone subsumes the spatial operation patterns of most convolutional architectures~\cite{yolov11}, including depthwise separable convolutions, residual blocks, etc.
The two workloads are:
\emph{instance segmentation} (Seg) with YOLO11m-seg on DAVIS~\cite{davis}, a video object segmentation benchmark containing 90 sequences (25--104 frames each) with varied camera motion, and
\emph{pose estimation} (Pose) with YOLO11m-pose on 3DPW~\cite{3dpw}, an outdoor human activity dataset from a handheld camera, containing 61 sequences (273--2178 frames each).
All inputs are resized to $1024\times1024$ and evaluated with official checkpoints without fine-tuning.
Because the datasets are distributed as image sequences, we encode each with x264 in all-P-frame mode and extract block-level MVs at the standard $16\times16$ macroblock granularity.
DAVIS exhibits substantially stronger motion than 3DPW (MV std 23.5 vs.\ 10.7\,px), making it a more challenging setting for cache reuse; together, the two datasets cover complementary regimes of rapid motion and long-term stability.
Table~\ref{tab:workload-profile} summarizes the two workloads.

\begin{table}[t]
\caption{Workload profile.}
\label{tab:workload-profile}
\centering
\begin{tabular}{lcc}
\toprule
 & Seg (YOLO11m-seg) & Pose (YOLO11m-pose) \\
\midrule
Dataset & DAVIS & 3DPW \\
Input resolution & $1024\times1024$ & $1024\times1024$ \\
Parameters (M) & 22.4 & 20.9 \\
Edge latency (ms) & $537.5\pm41.2$ & $446.8\pm16.4$ \\
Edge energy (J) & 7.61 & 6.86 \\
Server latency (ms) & $35.7\pm3.7$ & $27.6\pm3.1$ \\
MV std (px) & 23.5 & 10.7 \\
\bottomrule
\end{tabular}
\begin{tablenotes}
    \item[$*$] Edge and server latency are dense execution times. Edge energy is measured per frame via INA3221 power integration. MV std measures motion intensity.
\end{tablenotes}
\end{table}

\paragraph{Ground truth and metrics}
For DAVIS segmentation, we generate pseudo ground truth by running dense YOLO11x-seg on every frame, since the native annotations do not match the YOLO label set; all methods are compared against the same pseudo-GT, so the reported mIoU measures relative accuracy retention.
For 3DPW pose estimation, we use the dataset ground truth directly and report Object Keypoint Similarity (OKS).
End-to-end latency includes preprocessing, dispatch, transmission, inference, and cache maintenance.
Edge energy is measured by integrating board-level power over each frame interval; the reported value is the per-frame average over the mixed online execution path.

\paragraph{Baselines}
We compare FluxShard with the following four baselines.
\begin{itemize}
    \item \textbf{Offload} sends every full frame to the cloud server for dense inference, representing the pure-offload upper bound on network cost.
    \item \textbf{COACH}~\cite{coach} performs a whole-frame SSIM check: if similarity is high the entire output is reused; otherwise, the full frame is recomputed or transmitted. The frame is quantized to 1/4 size during transmission for bandwidth reduction.
    \item \textbf{DeltaCNN}~\cite{deltacnn} propagates only the per-pixel difference between consecutive frames through the network, skipping positions whose activation change is below a fixed threshold; it maintains the feature cache in a fixed coordinate system without motion compensation.
    \item \textbf{MotionDeltaCNN}~\cite{motiondeltacnn} (referred to as M-DeltaCNN) extends DeltaCNN by estimating a single global homography from the frame pair and shifting the entire cache accordingly before computing deltas; it still treats the cache as a rigid whole, so heterogeneous local motion is not handled at the shard level.
\end{itemize}
DeltaCNN uses the original open-sourced CUDA implementation~\cite{deltacnn}; M-DeltaCNN is not open-sourced and is therefore re-implemented on our sparse backend, making its comparison with FluxShard the most direct.
All baselines (except Offload) share the same profiling-driven dispatch logic as FluxShard to isolate reuse semantics from transport or scheduling differences.
Statistics exclude the first (initialization) frame of each sequence.

\subsection{End-to-End Performance}\label{sec:eval-e2e}

\begin{figure}[t]
    \centering
    \includegraphics[width=\columnwidth]{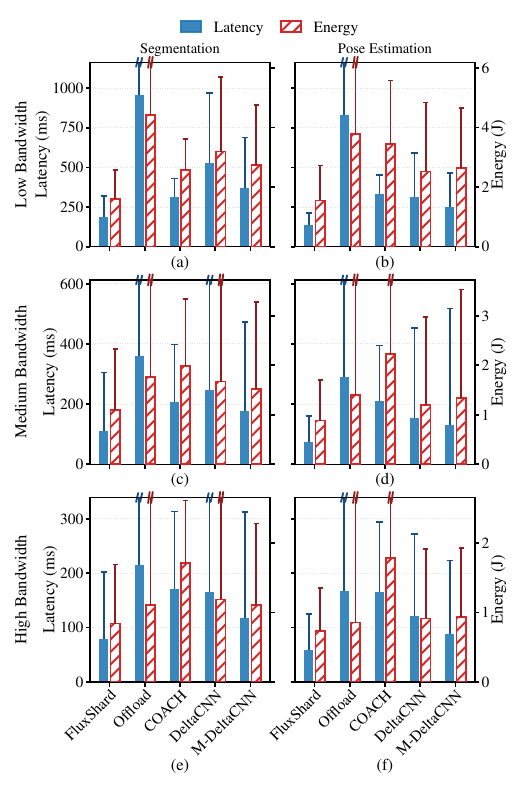}
    \caption{End-to-end latency (solid bars, left axis) and edge energy (hatched bars, right axis) under three bandwidth tiers.
    Within each row, the two panels share the same axis range for direct comparison.
    Error bars show one standard deviation; break markers indicate clipped bars.}
    \label{fig:e2e-latency-energy}
\end{figure}

\paragraph{Latency and energy under dynamic bandwidth}
Figure~\ref{fig:e2e-latency-energy} compares the realized online latency and edge energy across bandwidth tiers.
FluxShard achieves the lowest latency and energy across both workloads and all tiers.
Motion-aligned cache remapping keeps the recomputation set small, which simultaneously reduces the transmitted payload and the computation on both endpoints; the dispatcher then routes the smaller workload to the endpoint with lower latency.

For segmentation, FluxShard delivers 33.0--80.4\% latency reduction and 23.7--64.0\% energy savings across the three tiers; under low bandwidth, latency drops from 955\,ms (Offload) to 187\,ms and energy from 4.43\,J to 1.59\,J.
For pose estimation, the gains persist: 32.6--83.8\% latency reduction and 14.9--59.2\% energy savings, with latency dropped from 832\,ms to 134\,ms under low bandwidth.
The advantage is the widest under low bandwidth, where transmission dominates and FluxShard's small recomputation payload cuts the bottleneck directly.

Among the reuse baselines, the relative ranking between COACH and the DeltaCNN variants shifts with bandwidth. 
Under low bandwidth, COACH's quantization compresses the full-frame payload enough to beat DeltaCNN on segmentation (312\,ms vs.\ 525\,ms), because both DeltaCNN variants lose sparsity under strong motion and their recomputation payloads saturate the link.
Under medium and high bandwidth, M-DeltaCNN's motion-compensated sparsity helps it overtake COACH under both workloads, while DeltaCNN sometimes still trails COACH on segmentation where camera motion is more diverse.
FluxShard's advantage holds across all settings because its shard-level MV alignment preserves reuse under motion that invalidates fixed-coordinate or whole-frame approaches.

\paragraph{Accuracy}
Table~\ref{tab:e2e-accuracy} reports accuracy across all settings.
Our profiling-based configuration targets at most 3\% accuracy loss relative to dense offload (the constraint $\bar{A} \ge \alpha \bar{A}^*$ in Eq.~\eqref{eq:objective}).
FluxShard stays within 2.0--2.7\% across both workloads and all tiers, meeting the budget with margin.
DeltaCNN and M-DeltaCNN lose 1.6--2.9\%; DeltaCNN retains slightly higher accuracy on segmentation (1.6--1.7\% vs.\ FluxShard's 2.0--2.7\%) because its fixed-coordinate cache avoids the additional approximation introduced by MV alignment and truncation, at the cost of substantially higher latency.
COACH incurs the largest drop (5.4--7.2\%): whole-frame similarity does not ensure accurate dense predictions, and quantization of the transmitted frames further compounds the loss.

\begin{table}[t]
\caption{End-to-end accuracy under trace-driven bandwidth replay.}
\label{tab:e2e-accuracy}
\centering
\setlength{\tabcolsep}{2.5pt}
\begin{tabular}{clccc}
\toprule
Workload & Method & Low & Medium & High \\
\midrule
\multirow{5}{*}{\shortstack{Seg\\(mIoU)}} & Offload & 69.15 & 69.15 & 69.15 \\
 & FluxShard & 67.75 (-2.02\%) & 67.71 (-2.08\%) & 67.30 (-2.68\%) \\
 & COACH & 65.42 (-5.40\%) & 65.18 (-5.74\%) & 64.87 (-6.19\%) \\
 & DeltaCNN & 68.06 (-1.58\%) & 68.07 (-1.57\%) & 67.99 (-1.68\%) \\
 & M-DeltaCNN & 67.25 (-2.75\%) & 67.19 (-2.84\%) & 67.17 (-2.86\%) \\
\midrule
\multirow{5}{*}{\shortstack{Pose\\(OKS)}} & Offload & 70.95 & 70.95 & 70.95 \\
 & FluxShard & 69.21 (-2.45\%) & 69.13 (-2.56\%) & 69.02 (-2.72\%) \\
 & COACH & 67.14 (-5.37\%) & 66.19 (-6.71\%) & 65.85 (-7.19\%) \\
 & DeltaCNN & 69.00 (-2.75\%) & 69.02 (-2.72\%) & 69.00 (-2.75\%) \\
 & M-DeltaCNN & 69.22 (-2.44\%) & 69.03 (-2.71\%) & 68.93 (-2.84\%) \\
\bottomrule
\end{tabular}
\begin{tablenotes}
    \item[$*$] Parenthesized values show the percentage accuracy drop relative to dense offload.
\end{tablenotes}
\end{table}

\subsection{Microbenchmark}\label{sec:eval-micro}

To understand where the end-to-end gains come from, we examine per-frame component statistics under medium bandwidth.

\paragraph{Ratio breakdown}
Table~\ref{tab:microbench-ratio} reports the transmission ratio (proportion of full frame transmitted), the computation ratio (proportion of dense FLOPs executed), and the cloud dispatch ratio (proportion of frames routed to cloud) under medium bandwidth for both workloads.
Across all methods, segmentation transmission and computation ratios are consistently higher than pose, reflecting the stronger motion in DAVIS (MV std 23.5 vs.\ 10.7\,px, Table~\ref{tab:workload-profile}): more motion invalidates more cached positions, increasing both transmitted payload and computation.

FluxShard transmits only 8.9--17.1\% of the full frame and executes 40.7--47.9\% of dense FLOPs, compared with 28.9--56.9\% and 54.6--66.8\% for the DeltaCNN variants.
COACH transmits 24\% owing to $4\times$ quantization, but computes over 96\% of dense FLOPs because its whole-frame SSIM check rarely triggers reuse on dynamic video.
In contrast, FluxShard achieves a lower transmission ratio while halfing computation, with only 2.0--2.7\% accuracy loss (Table~\ref{tab:e2e-accuracy}).
The cloud dispatch ratio is high for all methods (93--100\%) due to the large compute gap between the two endpoints.
Peak GPU memory stays below 2.3\,GB for all methods; FluxShard's feature caches add 72--136\,MB over dense inference, comparable to DeltaCNN.

\begin{table}[t]
\caption{Ratio statistics of different methods under medium bandwidth.}
\label{tab:microbench-ratio}
\centering
\setlength{\tabcolsep}{1.5pt}
\begin{tabular}{lcccccc}
\toprule
\multirow{2}{*}{Method} & \multicolumn{3}{c}{Seg} & \multicolumn{3}{c}{Pose} \\
\cmidrule(lr){2-4} \cmidrule(lr){5-7}
& Tx (\%) & Comp (\%) & Cloud (\%) & Tx (\%) & Comp (\%) & Cloud (\%) \\
\midrule
FluxShard & 17.1 & 47.9 & 98.1 & 8.9 & 40.7 & 95.6 \\
Offload & 100.0 & 100.0 & 100.0 & 100.0 & 100.0 & 100.0 \\
COACH & 24.0 & 96.0 & 97.5 & 24.1 & 96.5 & 93.4 \\
DeltaCNN & 56.9 & 66.8 & 98.4 & 31.7 & 54.6 & 96.4 \\
M-DeltaCNN & 45.5 & 65.7 & 98.0 & 28.9 & 58.5 & 93.7 \\
\bottomrule
\end{tabular}
\begin{tablenotes}
    \item[$*$] Transmission ratio (Tx) is the fraction of full frame transmitted. Computation ratio (Comp) is the fraction of dense FLOPs executed. Cloud dispatch ratio (Cloud) is the fraction of frames routed to cloud.
\end{tablenotes}
\end{table}

\begin{figure}[t]
    \centering
    \includegraphics[width=\columnwidth]{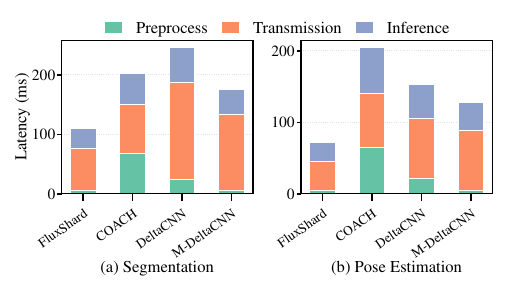}
    \caption{Mean per-frame latency breakdown under medium bandwidth.
    Preprocessing includes both edge-side and server-side preprocessing.
    DeltaCNN uses its open-sourced CUDA engine; M-DeltaCNN uses our backend.}
    \label{fig:microbench-breakdown}
\end{figure}

\paragraph{Latency breakdown}
Figure~\ref{fig:microbench-breakdown} decomposes mean per-frame latency into preprocessing, transmission, and inference.
FluxShard has the smallest total bar on both workloads: its low transmission and computation ratios jointly shrink both network and compute components.
Despite similar computation ratios (Table~\ref{tab:microbench-ratio}), DeltaCNN and M-DeltaCNN show different inference latencies due to their different execution backends; we explain this gap below.

\begin{figure}[t]
    \centering
    \includegraphics[width=\columnwidth]{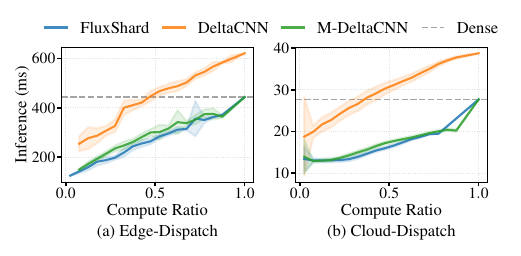}
    \caption{Inference latency vs.\ compute ratio for pose estimation under medium bandwidth, split by dispatch path.
    Dashed lines show dense inference latency on each endpoint.
    FluxShard and M-DeltaCNN share our sparse execution backend (overlapping trends); DeltaCNN uses its original open-sourced CUDA implementation (different absolute level but similar slope).}
    \label{fig:microbench-trend}
\end{figure}

\paragraph{Sparsity-to-latency relationship}
Figure~\ref{fig:microbench-trend} plots inference latency against compute ratio for pose estimation under medium bandwidth, split by dispatch path.
FluxShard and M-DeltaCNN share our sparse backend, so their trends overlap; the latency difference comes from the compute ratio each achieves.
DeltaCNN's original engine follows a similar near-linear trend at a different absolute level, explaining the latency gap in Figure~\ref{fig:microbench-breakdown}.
The near-linear trend confirms that FluxShard's low compute ratio translates into proportionally low inference latency, which is particularly important under high bandwidth where computation rather than transmission is the bottleneck.

\subsection{Ablation Study}\label{sec:eval-ablation}

Table~\ref{tab:ablation-rfap} ablates key FluxShard components under medium bandwidth.
The default FluxShard uses the compacted input-level RFAP check; we compare against four variants.

\paragraph{RFAP variants}
We compare three RFAP configurations: disabled, compacted input-level (default), and per-layer.
Disabling RFAP entirely (w/o RFAP) yields the lowest compute ratio (Seg 45.0\%, Pose 38.7\%) because no positions are forcibly invalidated, but accuracy degrades substantially (Seg $-3.29\%$, Pose $-5.13\%$), with Pose exceeding the 3\% budget.
Without RFAP, positions where the input and output MV correspondences disagree are silently reused, and the resulting errors accumulate through the network.
The impact is smaller on Seg because its stronger motion already produces a larger input-level active set (Seg tx ratio 17.1\% vs.\ Pose 8.9\%, Table~\ref{tab:microbench-ratio}), which is more likely to overlap with the positions RFAP would have invalidated.

Per-layer RFAP checks MV consistency at every layer independently, achieving the highest accuracy among the three variants (Seg $-1.75\%$, Pose $-1.87\%$) but inflating the compute ratio to Seg 61.9\% and Pose 48.8\%, with latency also rising to Seg 117.1\,ms and Pose 75.6\,ms.
The redundant per-layer checks force recomputation at deeper-layer positions whose actual discrepancy magnitude is small enough for truncation to handle.

The compacted input-level check (default) strikes the best trade-off: it captures all structurally inconsistent positions at the input level where MV heterogeneity is explicit, and lets the profiled thresholds $\{\tau_l\}$ absorb small discrepancies at deeper layers.
This keeps the compute ratio at Seg 47.9\% and Pose 40.7\% with accuracy drop of Seg $-2.08\%$ and Pose $-2.56\%$.

\paragraph{w/o offload}
Restricting all inference to the edge device (w/o offload) yields Seg $-0.30\%$ and Pose $-2.94\%$ accuracy drop.
Latency, however, increases dramatically: Seg rises to 433.9\,ms and Pose to 252.4\,ms, a 3.4--3.9$\times$ slowdown, confirming that sparse reuse alone cannot compensate for the loss of cloud compute capacity.

\paragraph{w/o sparse}
Disabling sparse computation while keeping the original transmission logic (w/o sparse) yields Seg $-1.90\%$ and Pose $-0.23\%$ accuracy drop.
Latency rises to Seg 124.9\,ms and Pose 92.0\,ms, a 1.1--1.2$\times$ increase over the full system, because the server executes dense inference on every offloaded frame, instead of reusing cached features.

\paragraph{w/o remap}
Removing MV-guided cache remapping (w/o remap) causes the feature cache to drift out of alignment with the current frame over time.
As the accumulated displacement grows, RFAP and truncation detect an increasing number of inconsistent positions and force recomputation, driving compute ratio to Seg 77.1\% and Pose 77.3\%.
The same drift degrades the sparsity estimate used by the dispatcher, increasing the transmitted payload per frame; together, latency climbs to Seg 226.4\,ms and Pose 198.1\,ms---roughly $2\times$ the default and exceeding even the w/o sparse baseline.
Accuracy manifests an asymmetric effect: Pose degrades to $-10.61\%$ because keypoint localization is susceptible to coordinate-system misalignment, while Seg shows only $-1.07\%$, marginally better than the default ($-2.08\%$).
At a compute ratio of 77.1\%, the system recomputes the vast majority of positions, leaving nearly no room for reuse errors to manifest; the near-dense execution incidentally eliminates the small approximation errors that the default's aggressive reuse introduces.
The marginal accuracy difference (1\%) is well within the noise floor of the 3\% budget and comes at the cost of $2\times$ latency, confirming that cache remapping is essential for efficiency.

\begin{table}[t]
\caption{Ablation study under medium bandwidth.}
\label{tab:ablation-rfap}
\centering
\setlength{\tabcolsep}{2.5pt}
\begin{tabular}{clccc}
\toprule
Workload & Variant & Acc (\%) & Comp (\%) & Lat (ms) \\
\midrule
\multirow{6}{*}{\shortstack{Seg\\(mIoU)}} & FluxShard & 67.71 (-2.08\%) & 47.9 & 111.7 \\
 & w/o RFAP & 66.87 (-3.29\%) & 45.0 & 115.0 \\
 & Per-layer RFAP & 67.94 (-1.75\%) & 61.9 & 117.1 \\
 & w/o offload & 68.94 (-0.30\%) & 45.4 & 433.9 \\
 & w/o sparse & 67.84 (-1.90\%) & 100.0 & 124.9 \\
 & w/o remap & 68.41 (-1.07\%) & 77.1 & 226.4 \\
\midrule
\multirow{6}{*}{\shortstack{Pose\\(OKS)}} & FluxShard & 69.13 (-2.56\%) & 40.7 & 74.1 \\
 & w/o RFAP & 67.31 (-5.13\%) & 38.7 & 77.6 \\
 & Per-layer RFAP & 69.62 (-1.87\%) & 48.8 & 75.6 \\
 & w/o offload & 68.87 (-2.94\%) & 41.1 & 252.4 \\
 & w/o sparse & 70.79 (-0.23\%) & 100.0 & 92.0 \\
 & w/o remap & 63.42 (-10.61\%) & 77.3 & 198.1 \\
\bottomrule
\end{tabular}
\begin{tablenotes}
    \item[$*$] FluxShard uses compacted input-level RFAP as the default configuration. Parenthesized values show the percentage accuracy drop relative to dense offload.
\end{tablenotes}
\end{table}

\subsection{Scalability}\label{sec:eval-scalability}

To evaluate how FluxShard scales when multiple edge devices share the same cloud server, we run 1, 2, and 3 concurrent edges under medium bandwidth.
Each additional edge runs as an independent client process on a separate Jetson Xavier NX; all clients share the same cloud server and the same \texttt{tc}-shaped uplink.
Because each client maintains its own cache state and accuracy is determined by the per-client reuse policy (which is unchanged), we focus on latency and energy.

\begin{figure}[t]
    \centering
    \includegraphics[width=\columnwidth]{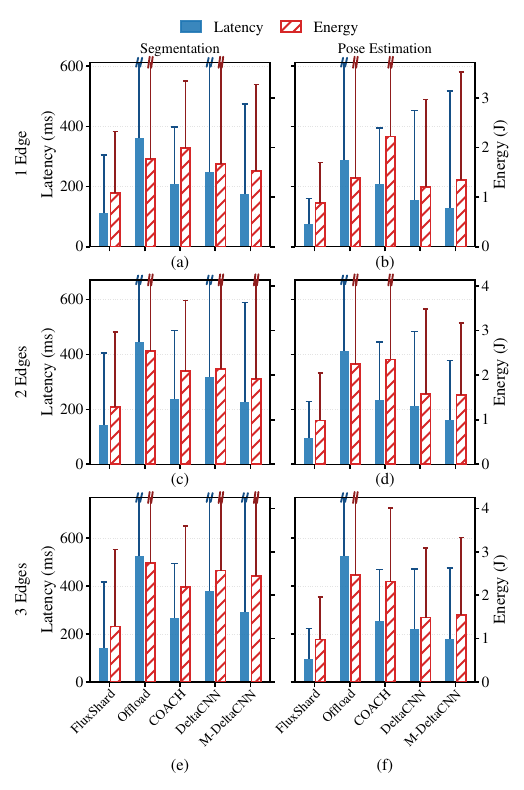}
    \caption{End-to-end latency (solid bars, left axis) and edge energy (hatched bars, right axis) under 1, 2, and 3 concurrent edge devices under medium bandwidth.}
    \label{fig:scalability}
\end{figure}

Figure~\ref{fig:scalability} compares latency and energy as the number of active edges increases.
FluxShard's latency rises by only 28\% from 1 to 3 edges on pose estimation (74.1\,ms to 95.2\,ms), while Offload degrades by 82\% (289.8\,ms to 526.6\,ms); segmentation follows a similar pattern.
Among the reuse baselines, M-DeltaCNN and DeltaCNN rise by 37\% and 42\% respectively on pose, both larger than FluxShard.
FluxShard's advantage widens with more clients because its small transmission and recomputation payload imposes less load on the shared server.
Edge energy follows the same trend: FluxShard rises by only 11\% on pose, while Offload nearly doubles.

\subsection{Sensitivity}\label{sec:eval-sensitivity}

Table~\ref{tab:sensitivity} reports pose estimation under medium bandwidth as we vary the accuracy budget $\alpha$ and the budget split ratio $r$ reserved for $\tau_0$ in Eq.~\eqref{eq:truncation-budget-split}.
We omit segmentation as it exhibits similar trends.
All configurations keep accuracy within their respective budgets.
Tightening $\alpha$ from 0.03 to 0.01 raises the compute ratio from 40.7\% to 52.0\%, and transmission ratio from 8.9\% to 12.4\%, increasing latency by 28\%, while relaxing $\alpha$ to 0.05 reduces all these metrics.
Increasing $r$ from 0.50 to 0.90 allocates more budget to the dispatch layer.
A larger $\tau_0$ admits more input positions as reusable, reducing the transmission ratio from 9.0\% to 6.9\%; the remaining DNN layers receive less budget, so their thresholds tighten and the compute ratio rises from 39.0\% to 45.1\%.
Under medium bandwidth where transmission is the bottleneck, the net effect is a latency reduction from 75.3\,ms to 71.5\,ms with accuracy stable around $-2.6\%$, illustrating that tuning the budget split to match the deployment bottleneck can yield meaningful gains.

\subsection{Discussion}\label{sec:eval-discussion}

When the camera is static, per-block MVs are near zero and cache remapping becomes a no-op; FluxShard degrades gracefully to behavior similar to DeltaCNN.
Under extreme motion such as scene cuts, the reuse ratio approaches zero and FluxShard falls back to full recomputation; this is the expected lower bound, as no temporal reuse method can exploit non-existent temporal redundancy.

\paragraph{MV quality}
Codec MVs are the output of block matching during encoding, not optical flow; in texture-flat regions or under occlusion, the encoder may select a displacement that does not reflect the true motion.
Inaccurate MVs do not cause silent accuracy degradation, however, because FluxShard's correctness relies on two independent checks: RFAP detects structural inconsistency within the receptive field, and the content-change threshold $\tau_0$ detects value mismatches after MV alignment.
If an MV points to the wrong source, the aligned pixel value will differ from the current frame, and the position will enter the recomputation set and be recomputed.
The only consequence of inaccurate MVs is a lower reuse ratio, which increases latency but never compromises accuracy.

FluxShard's mechanisms are defined for convolutional architectures with fixed receptive fields and spatial strides.
RFAP's two conditions depend only on receptive field geometry and layer strides, properties shared by all standard convolutional layers (regular, depthwise separable, dilated, and grouped convolutions) regardless of the surrounding network topology.
Cache remapping relies solely on the MV field and backward warping, which operate on spatial grids independent of layer type.
Because YOLO11's backbone subsumes the spatial operation patterns of most CNN-based detection and segmentation architectures~\cite{yolov11}, the evaluation spans sufficient structural diversity.
Attention-based layers, however, aggregate information from data-dependent positions rather than a fixed spatial neighborhood, so Condition~1 (intra-receptive-field uniformity) does not directly apply.
Extending RFAP to architectures with dynamic receptive fields is an open direction for future work.

\begin{table}[t]
\caption{Sensitivity to different configurations\\(Pose, medium bandwidth).}
\label{tab:sensitivity}
\centering
\setlength{\tabcolsep}{2.5pt}
\begin{tabular}{ccccccc}
\toprule
\multicolumn{2}{c}{Config} & Accuracy & \multicolumn{2}{c}{Ratio (\%)} & Lat & Energy \\
\cmidrule(lr){1-2} \cmidrule(lr){3-3} \cmidrule(lr){4-5} \cmidrule(lr){6-6} \cmidrule(lr){7-7}
$\alpha$ & $r$ & OKS (\%) & Tx & Comp & (ms) & (mJ) \\
\midrule
0.03 & 0.67 & 69.13 (-2.58\%) & 8.9 & 40.7 & 74.8 & 891 \\
0.03 & 0.50 & 69.01 (-2.75\%) & 9.0 & 39.0 & 75.3 & 899 \\
0.03 & 0.90 & 69.08 (-2.65\%) & 6.9 & 45.1 & 71.5 & 874 \\
0.01 & 0.67 & 70.31 (-0.92\%) & 12.4 & 52.0 & 96.1 & 1111 \\
0.05 & 0.67 & 67.55 (-4.81\%) & 6.9 & 34.0 & 72.8 & 873 \\
\bottomrule
\end{tabular}
\begin{tablenotes}
    \item[$*$] Default: $\alpha{=}0.03$, $r{=}0.67$. $\alpha$: accuracy retention budget in Eq.~\eqref{eq:objective}. $r$: fraction of the accuracy budget reserved for $\tau_0$ (Eq.~\eqref{eq:truncation-budget-split}). Parenthesized values show accuracy change relative to dense offload (OKS\,70.96).
\end{tablenotes}
\end{table}

\section{Related Work}\label{sec:related_work}

\textbf{Edge intelligence workloads.}
Edge intelligence deploys DNN inference on resource-constrained devices, demanding real-time perception and decision-making at the point of data generation.
Representative workloads span a wide range of visual tasks: UAV delivery systems rely on on-board visual positioning for autonomous navigation~\cite{s23218711,nguyen_person_2025,keepedge_tmc23}; mobile robots use visual SLAM to build maps and localize in real time~\cite{wu_helpful_2024,han_fetchbench_2024,visual_slam_tmc24}; augmented-reality applications require low-latency scene understanding to overlay virtual content~\cite{zhang_vdo-slam_2021,ifresher_tmc25}; edge video analytics pipelines perform continuous object detection~\cite{jalali_robust_2025,realtime_detection_tmc25} and multi-object tracking~\cite{fasttuner_tmc25,mystique_tmc25} under strict latency budgets.
These tasks all require dense predictions on every frame of a continuous video stream, where both latency and accuracy are critical.

Consecutive frames share substantial visual content, and this redundancy extends to intermediate feature maps; caching and reusing these features across frames reduces both computation and transmission. Existing approaches fall into two broad categories: pipeline-level caching and delta-based sparse inference.

\textbf{Pipeline-level caching.}
Methods, such as SPINN~\cite{spinn} and COACH~\cite{coach}, cache intermediate feature maps or predictions and reuse them when successive inputs are globally similar, making a \emph{binary, whole-input} reuse decision. This coarse granularity suits classification but cannot approximate the spatially dense outputs required by segmentation or detection.

\textbf{Delta-based sparse inference.}
A second line of work, including RRM~\cite{rrm}, CBinfer~\cite{cbinfer}, Skip-Convolution~\cite{skip-conv}, and DeltaCNN~\cite{deltacnn}, maintains a pixel-level reference cache and propagates only the \emph{difference} (delta) between the current frame and the reference through the network, computing only at affected output locations and reusing cached values elsewhere. Among these, DeltaCNN represents the most complete realization: it achieves end-to-end sparse propagation with truncation buffers that prevent error accumulation, enabling unbounded-length sequences without dense resets and pushing the efficiency of static-camera settings to its practical limit. MotionDeltaCNN~\cite{motiondeltacnn} extends DeltaCNN to moving cameras by warping the cache with a single global homography before computing the delta.

\textbf{Shared limitation.}
Across both categories, the cache is managed in a fixed or globally shifted coordinate system without per-region motion handling. Real-world mobile video, however, exhibits spatially non-uniform motion: a camera pan shifts the background uniformly while foreground objects move independently, and depth discontinuities create parallax that no single warp can reconcile. 
The result is a \emph{granularity mismatch}: the cache granularity is the whole scene, but motion granularity is per-region. This mismatch forces existing methods to either waste computation re-deriving content that has merely shifted, or sacrifice correctness by reusing misaligned features.

Note that several other techniques also reduce the cost of video analytics, including ROI filtering~\cite{glimpse,reducto}, input resolution adaptation~\cite{chameleon,accmpeg}, frame sampling that skips inference on selected frames~\cite{glimpse,reducto}, and model compression via quantization or pruning~\cite{quantization,han_learning_2015}.
Split-point inference methods partition a DNN at an intermediate layer and transmit compressed features rather than raw pixels~\cite{neurosurgeon,distributed_dnn_tmc24}, while multi-exit architectures allow early termination to reduce latency~\cite{multiexit_tmc24}.
Edge-cloud scheduling and offloading strategies further optimize where and when to execute inference under dynamic network conditions~\cite{joint_dnn_tmc25,video_offloading_tmc24}.
These strategies are orthogonal to feature cache reuse: they reduce the cost of a \emph{single} frame by compressing its representation or optimizing its placement, whereas caching exploits \emph{temporal} redundancy across frames to avoid redundant computation and transmission altogether.
FluxShard can be composed with any of them; this work focuses exclusively on the cache reuse axis.
\section{Conclusion}\label{sec:conclusion}
We have presented FluxShard, a motion-aware edge-cloud video analytics system that uses codec-level motion vectors to manage feature cache reuse and recomputation at the granularity of individual motion regions, achieving low latency and high accuracy on resource-constrained edge devices.
In FluxShard, RFAP identifies additional positions that must be recomputed due to heterogeneous motion, recovering accuracy and resolving the receptive field inconsistency problem; MV-guided cache remapping keeps the feature cache closely aligned with the evolving scene, improving the reuse ratio and addressing the cache maintenance problem; a profiling-driven dispatcher routes the remaining sparse workload to the faster endpoint.
By enabling real-time, high-accuracy video analytics on resource-constrained mobile platforms, FluxShard can serve as an inference infrastructure for latency-sensitive edge intelligence applications, such as autonomous drones, embodied robots, and augmented-reality devices. As a potential future direction, we are looking forward to extending our method to improve the performance of various applications, such as large language
models~\cite{lin2024splitlora,fang2026hfedmoe,lin2024split,qu2025mobile,fang2024automated} and distributed learning
system~\cite{lin2025hierarchical,wei2025optimizing,fang2026nsc,lin2025hasfl,hu2024accelerating,fang2026aggregation,lyu2023optimal,lin2024adaptsfl}.

\bibliographystyle{IEEEtran}
\bibliography{reference}

\end{document}